\documentclass{ws-procs9x6}

\begin{document}

\title{Quantum phase transition and engineering 
in two-component BEC in optical lattices}

\author{Yong-Shi Wu}

\address{Department of Physics, Univesity of Utah \\
Salt Lake City, UT 84112, USA\\ 
E-mail: wu@physics.utah.edu}

\maketitle

\abstracts{
In this paper we review recent progress in studying 
quantum phase transitions in one- and two-component 
Bose-Einstein condensates (BEC) in optical lattices. 
These phase transitions involve the emergence and 
disappearance of quantum coherence over whole optical 
lattice and of linear superposition of macroscopic 
quantum states. The latter may provide new means to 
engineer and to manipulate novel macroscopic quantum 
states and novel coherent atomic beams for quantum 
information processing, quantum computing etc.}

\section{Prologue}

The celebration of Professor Chen-Ning Yang's Eightieth 
Anniversary has a particular meaning to Chinese 
physicists of my age. A whole generation of Chinese 
physicists like me were inspired to choose physics as 
a career mainly by the news about the Nobel Prize of 
Physics in 1957. (Later we became to understand that 
among the great contributions of Prof. Yang to phyiscs, 
parity violation is actually {\it not} the greatest!) 
The celebration brings me back to the time of my 
first contact with Prof. Yang. During the early 
years of the Cultural Revolution (1966-1971), 
China was completely isolated from the world. This 
was particularly bad for young physicists who had 
just graduated from college, like me. When Prof. 
Yang first visited Beijing in 1971, I was lucky 
to be in the audience of his lectures in Beijing 
University. I still clearly remember his two lectures, 
i.e. the integral definition of the Yang-Mills field 
and the Bethe ansatz in one-dimensional exactly solvable 
many-body systems, respectively. The two lectures had 
great impacts on my later career in theoretical physics. 
In the first ten years afterwards, I was mainly working 
on Yang-Mills theory. Since I came to Stony Brook
in 1981 from China upon Prof. Yang's invitation, I have 
got involved with quantum statistical mechanics of 
fractional statistics (first with anyons and later 
with exclusons) in lower dimensional systems. Now I 
firmly believe that Quantum Yang-Baxter Equations 
will play a significant role in the foundation of 
string/M theory. 

The time allocated for my talk does not allow me 
to detail the other impacts that Prof. Yang had 
on my career. Here before presenting a recent 
work of mine with collaborator, I would like to 
express my wishes for Prof. Yang's good health, 
longevity and continuing success in his new career.

By now it is clear that we are in the early stages
of an exciting second "quantum revolution", which 
is bringing us to the intersection of the microscopic 
quantum and the macroscopic classical worlds. This 
common meeting ground is rich in new physics and in 
new technology. The goal of the study is to achieve 
eventually the engineering and manipulation of atoms 
and molecules, either individually or collectively, 
entirely at the quantum level. Among others, BEC 
as a new form of matter in the quantum regime is 
a fascinating system full of promises for quantum
manipulation and engineering. Below I will describe
a recent work of mine with Guang-Hong Chen on 
quantum phase transitions and appearance of novel
macroscopic quantum states in two-component BEC in
optical lattices.

\section{BEC as a Macroscopic Quantum State}

Prof. Yang recently has summarized the three main 
melodies of the twentieth century physics: {\it 
symmetry, quantization and phases}\cite{yang1}. 
Of course, the theory of Yang-Mills fields has 
been one of the successful syntheses of the three 
themes. Bose-Einstein condensates (BEC) may be 
viewed as another example of such synthesis: BEC 
are a collective {\it quantum} state, originated 
from {\it totally symmetric} wave function with 
respect to permutations of identical atoms of 
integer spin, in which there is a {\it phase 
coherence} over macroscopic distances. 

Originally Einstein predicted the existence of 
BEC, based on the assumption of ideal gas (with 
interactions between atoms ignored). It was 
Yang and collaborators\cite{yang2} in late
1950's who successfully treated the more realistic 
case of gaseous BEC with weakly interacting atoms. 
The paper of Byers and Yang\cite{yang3} in 1961
clarified the importance of being a macroscopic 
quantum state with phase coherence for the flux 
quantization in superconductivity, a phenomenon 
that shares many features with BEC.    
 
The transition in a trap between ordinary Bose gas 
and BEC occurs at a finite critical temperature 
$T_c$. At $T>>T_c$, a dilute gas consisting of 
identical atoms of integer spin behaves like a 
classical ideal gas, since two neutral atoms most 
of the time do not interact with each other unless 
they collide. However, as temperature decreases, 
the quantum mechanical "thermal de Broglie 
wavelength" of the atoms increases. At $T=T_c$ the 
thermal de Broglie wavelength becomes of the same 
order as the inter-particle distance, the wave 
packets of the atoms overlap with each other 
significantly. The whole system becomes, in a 
sense, a giant matter wave with a coherent 
phase. The Bose statistics of atoms begins to 
take effect, and all the atoms want to be in the 
same quantum state, forming a condensate called 
BEC. Thus it is clear that the transition to 
BEC at $T_c$ is controlled by temperature or by 
thermal fluctuations, so it is a {\it thermal} 
phase transition. Below $T_c$ the system remains 
to be a BEC which, as F. London suggested, 
exhibits superfluidity.

\section{Quantum Phase Transitions for BEC in
Optical Lattice}

The case in an optical standing wave field, 
or an optical lattice, formed by a number of 
counter-propagating laser beams is very 
different. Even at $T=0$, a phase transition 
from superfluid BEC to an insulator can happen 
by changing, say, the optical intensity. This 
is because the atoms, though neutral, interact 
with the optical standing wave through their 
induced electric dipole moment (the ac Stark 
effect). When the intensity of the laser beams 
is low, namely the optical period potential 
is weak, the atoms behave like Bloch waves, 
extending over the whole lattice like a 
superfluid. Let us tune the intensity of the 
laser beams to increase the strength of the 
optical lattice, then atoms tend to stay near 
the bottom of the valleys, i.e. the sites 
of the optical lattice, with a decreasing 
probability for tunneling between neighboring 
valleys. In this way, the BEC gets squeezed 
in optical lattice. If the (short-range) 
interactions between atoms are repulsive, then 
atoms do not feel happy to be in the same 
valley, and the system becomes {\it strongly 
correlated} due to the interactions between 
atoms in a narrow valley. 

Thus there is an interesting competition 
between inter-valley tunneling and intra-valley 
(or on-site) repulsive interactions: Tunneling 
makes atoms move from site to site, in favor of 
a superfluid BEC phase, at absolute zero, 
spreading over the whole lattice. On the other 
hand, the on-site repulsive interactions tend 
to forbid the motion to a neighboring site if 
that increases on-site potential energy. Therefore,
when the on-site energy is big enough, the atoms
do not move from site to site, and the system is 
in an insulating phase, called the Mott insulator,
to distinguish it from the usual band insulator.
Theoretically such a competition can be 
described\cite{zoller} by a lattice boson-Hubbard 
model\cite{fisher}. By increasing the intensity 
of the optical standing wave, one can reduce the 
tunneling probability and, at the same time, 
increase the on-site energy, because of the 
shrinking of the atomic wave function in a valley. 
In this way, we expect a phase transition from 
superfluid BEC to an insulating phase when the 
optical potential become sufficiently strong,
as indeed predicted by the boson Hubbard model. 
In the insulating phase, atoms are restricted
to individual valleys and can no longer hop
between different valleys. Therefore, the 
macroscopic quantum coherence across the 
system, that exists in the superfluid phase, 
is lost in the insulating phase. Namely 
accompanying the phase transition from BEC to 
(Mott) insulator is the loss of macroscopic 
quantum coherence, and the transition in the
opposite way is associated with the emergence
of macroscopic quantum coherence.

This phase transition is controlled by the 
intensity of laser beams that produce the 
optical lattice, not by temperature. It may 
happen even at absolute zero, driven by change
in quantum fluctuations. Such a phase transition
is called a {\it quantum} phase transition (QPT). 
In condensed matter theory, this is a hot topic 
in the frontier of strongly correlated systems. 
But normally in condensed matter systems it is 
hard to tune non-thermal parameters such as the 
on-site interactions. Also the presence of 
disorder either makes the observation of 
QPT very difficult or completely changes the 
behavior near transition. Free of these problems, 
the BEC in optical lattices is a clean system 
to experimentally observe and study QPT\cite{expt}.

\section{Two-component BEC in Optical Lattice}

For the purpose of quantum engineering, it is 
better to have more components in BEC to manipulate. 
In the case at hand, we consider atoms with same 
constituents but in, say, two different internal 
states, labeled as different species $A$ and $B$. 
Suppose the atomic polarizability, or the induced
electric dipole moments, for the two states are of 
{\it opposite} sign. Then atoms $A$ and $B$ will 
see different optical potentials with opposite sign; 
the valleys of the latter form two penetrating 
sub-lattices, labeled with the same letters $A$ 
and $B$, shifted relative to each other by a
quarter of the wave length of laser beams. 
Thus the system consists of two components 
with different species $A$ and $B$, each 
preferring to stay on its own sub-lattice. 

Such a two-component system was previously 
suggested in ref. \cite{zoller}. It was 
treated analytically the first time in the 
literature by Chen and me\cite{wu1}. We have 
been able to solve the ground state of a 
two-component boson-Hubbard model that 
describes this situation. And we predicted 
that by tuning the parameters of the optical 
standing wave and the density of atoms of 
each component, it is possible to make atoms 
in species $A$ or $B$ to be either in the 
BEC phase or in the Mott phase at will, and 
to make a transition between BEC and Mott 
phases for each component separately. A new 
piece of physics here is that there may be
interactions between atoms of species $A$ 
and $B$; we have shown that the pertinent 
interaction parameters can be exploited to 
adjust the phase boundaries. This gives us 
more flexibility for quantum manipulation.

\section{Engineering New Macroscopic Quantum 
States}

Furthermore, we suggest\cite{wu2} that a more 
interesting situation is to turn on an 
additional {\it Raman laser} to couple the two 
internal atomic states $A$ and $B$. The action 
of the coupling laser makes atoms continually 
convert from state $A$ to state $B$ and {\it vice 
versa}. After a conversion the atom will see a 
different optical potential, then it will move 
toward a neighboring site of the other sub-lattice. 
The frequent conversion and subsequent atomic 
motion between the sites of different sub-lattice 
couple the two components in the system together, 
and may develop phase coherence {\it between} 
the two components. Thus, one should expect to 
observe more interesting and perhaps more exotic 
phenomena, due to the {\it Raman-assisted 
tunneling between neighboring sub-lattice sites}.

As before, under appropriate conditions, atoms 
in internal state $A$ and those in $B$ both 
form superfluid BEC on sub-lattice $A$ and $B$, 
respectively. A preliminary analysis\cite{wu2}
done by Chen and me, based on a mean field 
theory, shows that due to Raman coupling, the 
order parameter in (or the macroscopic wave 
function of) the ground state of the system is 
actually a linear superposition of those 
for the two superfluid BEC on sub-lattice $A$ 
and $B$. In particular, there is a definite 
phase difference between the two superfluid 
components (on sub-lattice $A$ and $B$ 
respectively), implying an additional phase
coherence developed between the two superfluid 
components. The expression for the order
parameter of the system is formally similar to 
that for a superconducting Josephson junction.
However, physically this represents a new 
macroscopic quantum state, which has several 
characteristic features very different from
the case of the Josephson junction. 

Of course this state, on one hand, must share 
some similarities with the Josephson effect. 
On the other hand, unlike the case of the 
Josephson junction, the two phase-coherent 
superfluids are on {\it inter-penetrating} 
sub-lattices, rather than in separated 
spatial regions. Moreover, the atoms in the 
two coherent superfluids are in different 
internal states and, therefore, are {\it not} 
identical particles, while the Cooper pairs 
on different sides of a Josephson junction 
are identical to each other. Therefore, the 
constituent particles of the new quantum 
state are, in certain aspects, more like 
the $K^0_L$ particles or the neutrinos in 
neutrino oscillations in particle physics, 
which are linear superposition of states 
of {\it non-identical} particles. In other 
words, the Raman-assisted tunneling between 
sub-lattice $A$ and $B$ is accompanied by 
the conversion between different internal 
states, which has {\it no parallel} in the 
Josephson tunneling. More detailed study 
of the novel phase-coherent superposition 
of two {\it distinct} superfluids will be 
reported in the near future. 

In the original talk we raised the question 
of whether the new state could be viewed as 
a Schr\"{o}dinger cat. Now as of the present 
writing, we think that in a rigorous sense 
it is {\it not}, even though its macroscopic 
wave function (or order parameter) is indeed 
a linear superposition of those of two 
inter-penetrating macroscopic quantum states. 
However, the new macroscopic quantum state may 
be better viewed as {\it a condensate of 
Schr\"{o}dinger kittens}, i.e. a condensate 
of atoms in an internal state that is a linear 
superposition of two different internal states 
$A$ and $B$. When we turn off the trap and 
the optical lattice, we may expect to get a 
{\it novel coherent atomic beam}, in 
which each atom is a {Schr\"{o}dinger kitten} 
with relative amplitudes and phase {\it 
controllable} by adjusting optical parameters 
and atomic density. In this way, we expect 
that the two-component BEC in optical lattices 
provide new means for engineering and 
manipulating novel coherent atomic beams.

\section{Acknowledgement} I thank Prof. H.T. Nieh
and the Center of Advanced Study at Tsinghua
University in Beijing for inviting me to speak
at this Symposium and for their warm hospitality. 
Also I thank Dr. Guang-Hong Chen for pleasant
collaboration. This work was supported in part 
by the U.S. National Science Foundation under
Grant PHY-9907701.


\begin{thebibliography}{0}
\bibitem{yang1} For an earlier version, go to 
the website of Prof. C.N. Yang:  

http://insti.physics.sunysb.edu/Physics/faculty.htm\#yang

\noindent A recent version is presented by Prof. Yang at the 
International Conference of Theoretical Physics, UNESCO, 
Paris (July, 2002); see Proceedings. 
 
\bibitem{yang2} For an elegant summary of the results, 
see C.N. Yang, ``{\it Imperfect Bose System}'', {\it 
Physica} {\bf 26}, 549 (1960); and the references therein.  

\bibitem{yang3} N. Byers and C.N. Yang, {\it 
Phys. Rev. Lett.} {\bf 7}, 46 (1961).

\bibitem{zoller} D. Jaksch {\it et al.},  
{\it Phys. Rev. Lett.} {\bf 81}, 3108 (1961).

\bibitem{fisher} D.S. Fisher and M.P.A. Fisher,  
{\it Phys. Rev. Lett.} {\bf 61}, 1847 (1988).

\bibitem{expt} M. Greiner {\it et al.}, 
{\it Nature} {\bf 415}, 39 (2002).

\bibitem{wu1} G.H. Chen and Y.S. Wu, {\it Phys. 
Rev.} {\bf A67}, 013606 (2003).

\bibitem{wu2} G.H. Chen and Y.S. Wu, in 
preparation.





\end{thebibliography}
\end{document}